\documentclass[final,5p,times,twocolumn,number,sort&compress]{elsarticle}

\usepackage{amssymb}
\usepackage{lipsum}

\usepackage[bottom]{footmisc} %
\usepackage{graphicx}
\usepackage{lipsum}
\usepackage[hidelinks]{hyperref}
\usepackage{cleveref}
\usepackage{xcolor}

\usepackage{xfrac}
\usepackage{booktabs}
\usepackage{comment}

\journal{Physics Letters B}

\begin{document}

\crefname{figure}{Fig.}{Figs.}
\Crefname{figure}{Figure}{Figures}
\crefname{equation}{Eq.}{Eqs.}
\Crefname{equation}{Equation}{Equations}
\crefname{section}{Sec.}{Secs.}
\Crefname{section}{Section}{Sections}
\crefname{table}{Tab.}{Tabs.}
\Crefname{table}{Table}{Tables}
\creflabelformat{equation}{#2\textup{#1}#3}
\renewcommand{\crefrangeconjunction}{--}

\newcommand{\reftexts}{Ref.~}
\newcommand{\lips}[1]{\tilde{d} #1 \;}
\newcommand{\vecIV}[1]{#1} %
\newcommand{\vecIII}[1]{\vec{#1}} %
\newcommand{\vecII}[1]{\boldsymbol{#1}} %

\makeatletter

\newcommand{\coll}[2]{%
	$\coll@process{#1} + \coll@process{#2}$%
}

\newcommand{\collThree}[3]{%
	$\coll@process{#1} / \coll@process{#2} + \coll@process{#3}$%
}

\newcommand{\collFour}[4]{%
  $\coll@process{#1} / \coll@process{#2} / \coll@process{#3} + \coll@process{#4}$%
}

\newcommand{\coll@process}[1]{%
  \ifx#1A%
    #1%
  \else\ifx#1B%
    #1%
  \else\ifx#1x%
    #1%
  \else\ifx#1p%
    #1%
  \else\ifx#1d%
    #1%
  \else
    \coll@checkHeThree{#1}%
  \fi\fi\fi\fi\fi%
}

\newcommand{\coll@checkHeThree}[1]{%
  \ifnum\pdfstrcmp{#1}{He3}=0 %
    {}^3\mathrm{He}%
  \else
    \mathrm{#1}%
  \fi
}

\makeatother

\begin{frontmatter}

	\cortext[cor1]{Corresponding author. Email address: \href{frdcol002@myuct.ac.za}{frdcol002@myuct.ac.za} }

\title{A Unified Description of Small, Peripheral, and Large System\\Suppression Data from pQCD}

\author[first]{Coleridge Faraday}\corref{cor1}
\author[first]{W.\ A.\ Horowitz}
\affiliation[first]{organization={Department of Physics, University of Cape Town},
            addressline={Private Bag X3, Rondebosch 7701}, 
            country={South Africa}}

\begin{abstract}

We present quantitative predictions for the nuclear modification factor in both small and peripheral systems from a pQCD-based energy loss model that is constrained by light- and heavy-flavor suppression data from central heavy-ion collisions.
		We find nearly identical suppression for central \collThree{p}{d}{A} collisions as for peripheral \coll{A}{A} collisions, quantitatively consistent with the measured 20\% suppression of neutral pions produced in \coll{d}{Au} collisions by PHENIX, but dramatically inconsistent with the measured $20\%$ enhancement of charged hadrons produced in \coll{p}{Pb} collisions by ATLAS. 
	We demonstrate that this equivalence of central small system suppression and peripheral large system suppression is insensitive to the underlying energy loss model.

\end{abstract}

\end{frontmatter}

\section{Introduction}
\label{sec:intro}

There is currently overwhelming experimental evidence that quark-gluon plasma (QGP) is formed in central heavy-ion collisions at the Relativistic Heavy-Ion Collider (RHIC) and Large Hadron Collider (LHC). One set of important QGP signatures is comprised of low-momentum observables---such as elliptic flow, strangeness enhancement, and mean-$p_T$---which are related to the collective properties of the QGP, and are well-predicted by relativistic viscous hydrodynamics models \cite{Romatschke:2007mq, Song:2007ux, Schenke:2010nt}. Another set of QGP signatures considers the high-$p_T$ particles which are produced in heavy-ion collisions, where deviations from the proton-proton baseline are expected due to final state interactions of the particles with the medium. 
Final state suppression is typically measured by the nuclear modification factor, $R_{AB}^h \equiv (d N^{AB \to h} / d p_T) / (\left\langle N_{\text{coll}} \right\rangle d N^{pp \to h} / d p_T)$ for hadrons $h$ produced in the collision \coll{A}{B}, where $R_{AB} < 1$ indicates suppression.
A variety of partonic energy loss models \cite{Dainese:2004te, Schenke:2009gb, Wicks:2005gt, Horowitz:2012cf} based on perturbative Quantum Chromodynamics (pQCD) can, at least qualitatively, describe the $R_{AB}$ in central heavy-ion collisions \cite{PHENIX:2001hpc}.
These observables, among others, lead to the inevitable conclusion that a novel state of deconfined quark-gluon matter is formed in central heavy-ion collisions.

A pertinent open question in the field is whether small droplets of QGP are formed in small collision systems, including high-multiplicity \coll{p}{p} and \collFour{p}{d}{He3}{A} collisions \cite{Grosse-Oetringhaus:2024bwr}. This question was raised by the measurement of elliptic flow in high-multiplicity \coll{p}{A} collisions \cite{ATLAS:2012cix, ATLAS:2013jmi, ALICE:2014dwt, CMS:2015yux}, 
with subsequent evidence from strangeness enhancement \cite{ALICE:2015mpp, ALICE:2013wgn}, quarkonium suppression \cite{ALICE:2016sdt}, and higher order flow cumulants \cite{PHENIX:2018lia} implying that a QGP is formed in even the smallest of collision systems. Particularly compelling is the agreement between low-$p_T$ data and relativistic viscous hydrodynamics models \cite{Weller:2017tsr, Schenke:2020mbo}, similar to the heavy-ion case.

The experimental measurements in the high-$p_T$ sector do not provide as coherent a story as their low-$p_T$ counterparts. 
The ATLAS collaboration \cite{ATLAS:2022kqu} measures $R_{AB} \sim 1.2$ for $p_T \simeq 5\text{--}30 ~\mathrm{GeV}$ charged hadrons produced in $0\text{--}5\%$ centrality \coll{p}{Pb} collisions, which is not consistent with the expectation that final state energy loss leads to $R_{AB} < 1$. On the other hand, the PHENIX collaboration \cite{PHENIX:2023dxl} measures $R_{AB} \sim 0.75$ for $p_T \simeq 8\text{--}17 ~\mathrm{GeV}$  neutral pions produced in $0\text{--}5\%$ centrality \coll{d}{Au} collisions, which is qualitatively consistent with the picture of partons losing energy in a droplet of QGP. %
This apparent discrepancy may be due, at least in part, to biases \cite{ALICE:2014xsp,ALICE:2018ekf,ATLAS:2016xpn} in the Glauber Model (GLM) \cite{Miller:2007ri} based mapping between experimental measures of centrality and geometrical quantities, including the number of binary collisions $N_{\text{coll}}$.
	The ATLAS result \cite{ATLAS:2022kqu} is likely sensitive to this so-called \emph{centrality bias} as it employs the GLM-based mapping, while the PHENIX \cite{PHENIX:2023dxl} result---which is normalized by prompt, high-$p_T$ photons---is likely far less sensitive to these biases, because the GLM is not used in the measurement.

Due to the experimental difficulties present in measuring the $R_{AB}$, theoretical input is an invaluable and complementary tool in understanding the suppression pattern in small systems. In our previous works, we have discussed the phenomenological implications of including small system size corrections to both the radiative \cite{Faraday:2023mmx,Kolbe:2015rvk} and collisional energy loss \cite{Faraday:2024gzx,Wicks:2008zz}. In this work we perform a thorough statistical analysis to constrain our energy loss model to data from central heavy-ion collisions, and utilize the resulting large-system constrained model to make novel predictions for central small system and peripheral large system collisions. 
In order to isolate the effect of final state partonic energy loss in small, peripheral, and large colliding systems, we---similar to recent work by, e.g., JETSCAPE \cite{JETSCAPE:2024cqe}---neglect potential initial state effects such as nuclear PDFs (nPDFs) \cite{Eskola:2021nhw}, small-x evolution \cite{Armesto:2022mxy}, $k_T$ smearing \cite{Ke:2022gkq}, or color fluctuations \cite{Perepelitsa:2024eik} and final state hadronization effects such as coalescence \cite{Fries:2008hs} or in-medium fragmentation \cite{Zhao:2011cv}.

We will show that peripheral heavy-ion collisions form a plasma which is geometrically larger but cooler than that which may be formed in central small systems, such that one theoretically expects the same suppression in both systems. 
	The difference in size results in $\mathcal{O}(50)$ binary collisions for $60\text{--}70\%$ centrality heavy-ion collisions, compared to the $\mathcal{O}(10)$ binary collisions in $0\text{--}5\%$ centrality small system collisions, making peripheral heavy-ion collisions less sensitive to centrality bias than their central small system counterparts \cite{ALICE:2018ekf}.
	The combination of equal expected suppression and differing sensitivity to centrality bias in central small and peripheral large systems suggests that the lack of suppression measured in central \coll{p}{Pb} collisions may result from centrality bias.

\begin{figure}[!b]
	\includegraphics[width=\linewidth]{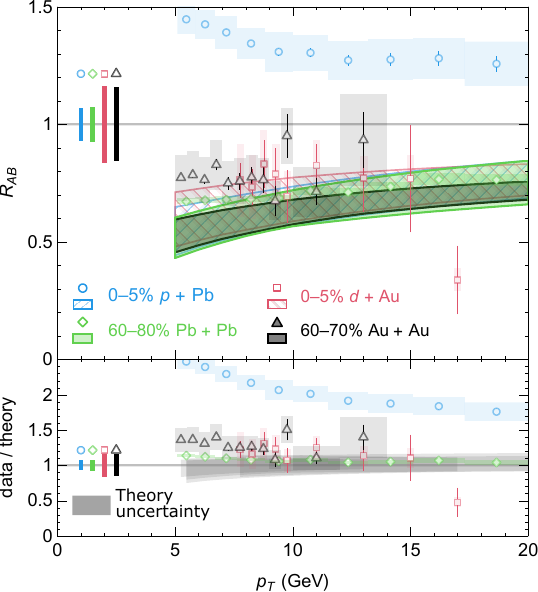}
	\caption{(top) Measured nuclear modification factor $R_{AB}$ (points) and large-system constrained model predictions (bands) as a function of $p_T$ for charged hadrons produced in $0\text{--}5\%$ centrality \coll{p}{Pb} and $60\text{--}80\%$ centrality \coll{Pb}{Pb} collisions at $\sqrt{s_{NN}} = 5.02 ~\mathrm{TeV}$ at LHC \cite{ATLAS:2022kqu}, as well as neutral pions produced in $0\text{--}5\%$ centrality \coll{d}{Au} \cite{PHENIX:2023dxl} and $60\text{--}70\%$ centrality \coll{Au}{Au} \cite{PHENIX:2012jha} collisions at $\sqrt{s_{NN}} = 0.2 ~\mathrm{TeV}$ at RHIC. The width of the model predictions corresponds to the theoretical uncertainty. (bottom) Measured $R_{AB}$ divided by theoretical prediction as a function of $p_T$ for the same systems as a function of $p_T$. Statistical uncertainties are represented by square brackets, systematic uncertainties by shaded squares, global normalization uncertainties by the solid rectangles at unity, and relative theoretical uncertainties by the gray shaded regions around unity. Points with statistical uncertainty $> 30 \%$ are excluded.}
	\label{fig:small_vs_peripheral}
\end{figure}

We show in \cref{fig:small_vs_peripheral} our predictions for light flavor hadron suppression in peripheral \coll{A}{A} and central \collThree{p}{d}{A} collisions at RHIC and LHC. Our predictions improve upon other similar high-$p_T$, small system results in the literature \cite{JETSCAPE:2024cqe,JETSCAPE:2024dgu,JETSCAPE:2021ehl,Huss:2020dwe,Huss:2020whe,Ke:2022gkq,Zakharov:2021uza} by: 1) removing the explicit large system size assumptions \cite{Kolbe:2015rvk,Wicks:2008zz,Faraday:2024gzx,Faraday:2023mmx} used by \emph{all} other energy loss models; and 2) constraining our model using \emph{only} large system data, and then making predictions for small systems with no additional tuning. We observe in \cref{fig:small_vs_peripheral} that the theoretical expectation from our large-system constrained energy loss model is in good agreement (within experimental and theoretical uncertainties) with the suppression data from central \coll{d}{Au} collisions, and peripheral \coll{Au}{Au} and \coll{Pb}{Pb} collisions, but is in stark disagreement with the enhancement observed in \coll{p}{Pb} collisions. Additionally, the theoretically expected $R_{AB}$ is extremely similar for central small and peripheral large systems at both RHIC and LHC.

We will further show that for \textit{any} of the popular energy loss models---based on weak or strong coupling, radiative or collisional energy loss, and thick or thin plasmas---to best describe peripheral heavy-ion data, one obtains strikingly similar theoretical estimates for suppression in central small systems as for peripheral large systems. This observation points towards a self-consistent description of energy loss in both central small and peripheral large system, resulting in equal suppression---consistent with \coll{d}{Au} \cite{PHENIX:2023dxl}, \coll{Pb}{Pb} \cite{ATLAS:2022kqu} and \coll{Au}{Au} \cite{PHENIX:2012jha} data---while in central \coll{p}{Pb} collisions, centrality bias may lead to an artificially enhanced $R_{AB}$ \cite{ATLAS:2022kqu}.

\section{Model description and statistical analysis}
\label{sec:results}

In this work we present results from our convolved radiative and collisional pQCD based energy loss model \cite{Faraday:2023mmx, Faraday:2024gzx} which receives small system size corrections to both the radiative \cite{Kolbe:2015rvk} and collisional energy loss \cite{Wicks:2008zz}. 
The radiative energy loss is the first order in opacity Djordjevic-Gyulassy-Levai-Vitev (DGLV) radiative energy loss \cite{Djordjevic:2003zk} which receives a short path length correction \cite{Kolbe:2015rvk}, and the collisional energy loss is derived in the framework of Hard Thermal Loops (HTL) \cite{Wicks:2008zz,Braaten:1991we}.
Partons propagate through a fluctuating collision geometry \cite{Schenke:2020mbo}, which is generated with IP-Glasma initial conditions and evolved according to longitudinal Bjorken expansion. We neglect running coupling effects in this work and treat the effective strong coupling constant $\alpha_s^{\text{eff.}}$ as the single free parameter in our model; however, a simple estimate for how running coupling may affect our conclusions is discussed later in this section. We map the initial temperature conditions $T(\mathbf{x})$ to distributions of effective lengths $L(\mathbf{x}_i, \hat{\mathbf{n}}) \equiv (1 / \langle T^3(\mathbf{x}_i \rangle) \int_0^\infty dz T^3(\mathbf{x}_i + z \hat{\mathbf{n}})$ where $\langle T^3 (\mathbf{x}_i \rangle \equiv \int d^2 \mathbf{x} T^6(\mathbf{x}) / \int d^2 \mathbf{x} T^3 (\mathbf{x})$, $\hat{\mathbf{n}}$ is the unit vector in the propagation direction of the hard parton, and $\mathbf{x}_i$ is the transverse location of the hard scatter. The temperature is calculated as $T(\mathbf{x}_i, \langle \tau \rangle) = (\langle T^3 \rangle)^{1 /3} (\tau_0 / \langle \tau \rangle)^{1 / 3}$, with the average time given by $\langle \tau \rangle = L / 2$.

We take into account three sources of theoretical uncertainty in our energy loss model. Firstly, there is an uncertainty in the crossover between HTL and vacuum propagators \cite{Wicks:2008zz, Faraday:2024gzx}. In our model this only affects the collisional energy loss; however, in principle, this uncertainty should also impact the radiative energy loss \cite{Wicks:2008zz}. This uncertainty is included in our model through two limiting derivations of the elastic energy loss: \emph{HTL-only} \cite{Wicks:2008zz} which uses only the HTL propagators and the derivation by Braaten and Thoma (\emph{BT}) \cite{Braaten:1991we} which uses HTL propagators for small momentum transfers and vacuum propagators for large momentum transfers.
Secondly, our previous work \cite{Faraday:2023mmx} found that the small system size corrections to the DGLV radiative energy loss becomes extremely large at high-momenta due to a breakdown in the large formation time approximation utilized in the DGLV radiative energy loss calculation. We found that while the breakdown of this assumption was most obvious once the short path length correction was included, the standard DGLV result also does not self-consistently respect the large formation time approximation at large momentum \cite{Faraday:2023mmx}. In lieu of a theoretical rederivation of the radiative energy loss result which removes the large formation time approximation, we advocated for restricting the phase space of gluon emission to reduce the effects of this breakdown \cite{Faraday:2023mmx,Faraday:2023uay}. 
A similar restriction is standard for enforcing consistency with the collinear approximation \cite{Horowitz:2009eb,Armesto:2011ht}.
As discussed in \cite{Horowitz:2009eb}, the collinear approximation imparts a large theoretical uncertainty on an extracted parameter such as $\hat{q}$ or $\alpha_s^{\text{eff.}}$ depending on the exact angle chosen for the cutoff; however, this uncertainty can largely be absorbed by a rescaling of the extracted parameter. In an analogous way, we restrict the phase space such that neither the large formation time nor the collinear approximations are ever violated. We estimate the theoretical uncertainty associated with this cutoff \cite{Faraday:2023mmx,Faraday:2023uay} by setting the upper bound on the transverse radiated gluon momentum as $k_{\text{max}} = \kappa \text{Min}[2 x E(1-x), \sqrt{2 x E} (\mu_D^2 + q^2)^{1 / 4}]$ and varying $\kappa$ in the range $[0.5, 2]$. 
Finally, we estimate the possible effect of running coupling on our results for central small and peripheral large systems, shown in \cref{fig:small_vs_peripheral}, using a simple procedure. The extracted coupling from large systems $\alpha_s^{\text{eff.}*}$ is modified in central small and peripheral large systems according to $\alpha_s^{\text{eff.}} \mapsto \alpha_s^{\text{eff.}}(T) =  (\alpha_s[Q(T)] / \alpha_s[Q(T_{\text{large}})]) \alpha_s^{\text{eff. }*}$. Here, $T$ is the temperature of the system under consideration, $T_{\text{large}}$ is the average temperature at which $\alpha_s^{\text{eff.}*}$ was extracted. The coupling is taken to run at $\alpha_s(Q) \propto [\log Q / \Lambda_{\text{QCD}}]^{-1}$ with $\Lambda_{\text{QCD}} = 0.2 ~\mathrm{GeV}$. We consider several common choices \cite{JETSCAPE:2024dgu,Xu:2014ica,Djordjevic:2013xoa,Peshier:2006ah} for the various scales $Q(T)$ at which the coupling runs; explicitly: $\in \{2 \pi T, \sqrt{ET}, \mu_D , \sqrt{4 E T}, \sqrt{2 E T}, E \}$. Each of the sources of theoretical uncertainty discussed here---crossover between vacuum and HTL propagators, the large formation time approximation, and running coupling---are $\mathcal{O}(1\text{--}10\%)$, leading to a total relative uncertainty, represented by the bands in \cref{fig:small_vs_peripheral}, of $\mathcal{O}(5\text{--}20\%)$.

We treat each of the variations in our energy loss model, which each have a specific $\kappa$ and collisional energy loss kernel (either BT or HTL), independently in the fitting procedure. This approach leads to different extracted values of $\alpha_s^{\text{eff.}}$ for each variation of the model, resulting in extracted values of $\alpha_s^{\text{eff.}}$ in the range $[0.3, 0.5]$ for the 14 different variations of our energy loss model in this analysis---an $\mathcal{O}(25\%)$ uncertainty on $\alpha_s^{\text{eff}}$. The uncertainties considered in this work are only a small subset of all uncertainties in theoretical energy loss calculations, including running coupling effects, higher order in $\alpha_s$ contributions, and treatment of energy loss in the pre-thermalization region. The already large uncertainty in the extracted value of $\alpha_s^{\text{eff.}}$, subject to further increases in uncertainty from the aforementioned contributions which are not yet treated, highlights the difficulty in attributing any physical meaning to the value of extracted parameters. For instance, the equivalent extraction of the jet quenching parameter $\hat{q} \propto \alpha_s^2$ would have an uncertainty of $\mathcal{O}(50\%)$, significantly larger than, for instance, the $\mathcal{O}(25\%)$ uncertainty on $\hat{q}$ at 90\% confidence interval in the most recent JETSCAPE extraction of $\hat{q}$ \cite{JETSCAPE:2024cqe}. Although it is difficult to associate a distribution or confidence interval with our theoretical uncertainty estimates, the magnitude of the variations in the extracted $\alpha_s$ warrants caution in drawing a correspondence between model parameters and physical parameters.

We extract $\alpha_s^{\text{eff.}}$ in our model by minimizing the effective $\chi^2$, following the procedure outlined in \cite{PHENIX:2008ove}. This analysis treats three forms of experimental uncertainty: \emph{type A}, which are uncorrelated; \emph{type B}, which have an unknown correlation matrix; and \emph{type C}, which are uncertainties common to each point, and are known to be fully correlated. Type A uncertainties include statistical uncertainties and uncorrelated systematic uncertainties; type B uncertainties include correlated systematic uncertainties; and type C uncertainties include global uncertainties related to the normalization of the $R_{AA}$, including those stemming from the proton-proton spectra and the number of binary collisions. Currently, the correlation matrix for the systematic uncertainties is not reported by experiments and so must be modelled. In this work, we consider three limiting hypotheses for the type B uncertainties: (I) the type B uncertainties are fully correlated and lead to all points shifting by the same fraction of the systematic uncertainty band on the point; (II) the type B uncertainties are fully correlated such that points at low and high momentum shift in opposite directions, and linearly interpolated in between; (III) the type B uncertainties are completely uncorrelated, and are therefore added in quadrature with the type A uncertainties. The I and II procedures were used in the PHENIX analysis \cite{PHENIX:2008ove}, and procedure III was added by us after we found that utilizing I and II only leads to an unstable extraction of $\alpha_s^{\text{eff.}}$ for high-$p_T$ pion data from the LHC. That is, the small statistical uncertainties in high-$p_T$ pion data from the LHC highlighted that the systematics are not well approximated by either of the I and II procedures, and variation of the systematic uncertainties on a point-by-point basis is needed for the model to reasonably describe the data.
The minimum $\chi^2$ obtained through procedures I, II, and III is then itself minimized to find the best fit for the model parameter $\alpha_s^{\text{eff.}}$. 
A more realistic model of the covariance matrix might allow data points which are close together to be more strongly correlated than those which are further apart, as described and implemented by the JETSCAPE collaboration \cite{JETSCAPE:2021ehl, JETSCAPE:2024cqe}. One may view our procedures as being limiting cases of this approach, with infinite correlation length (corresponding to I) and zero correlation length (corresponding to III). %

The experimental data used in the extraction of the $\alpha_s^{\text{eff.}}$ parameter were CMS \cite{CMS:2016xef}, ATLAS \cite{ATLAS:2022kqu}, and ALICE \cite{ALICE:2018vuu} charged hadron $R_{AA}$ measured in \coll{Pb}{Pb} collisions at $\sqrt{s_{NN}} = 5.02 ~\mathrm{TeV}$; CMS \cite{CMS:2017qjw} and ALICE \cite{ALICE:2018lyv} prompt $D$ meson $R_{AA}$ measured in \coll{Pb}{Pb} collisions at $\sqrt{s_{NN}} = 5.02 ~\mathrm{TeV}$; PHENIX \cite{PHENIX:2008saf,PHENIX:2012jha}  $\pi^0$ $R_{AA}$ measured in \coll{Au}{Au} collisions at $\sqrt{s_{NN}} = 0.2 ~\mathrm{TeV}$; and STAR \cite{STAR:2018zdy} $D$ meson $R_{AA}$ measured in \coll{Au}{Au} collisions at $\sqrt{s_{NN}} = 0.2 ~\mathrm{TeV}$. Data within the $0\text{--}50\%$ centrality-range and the $8\text{--}50 ~\mathrm{GeV}$ $p_T$-range was used in the $\alpha_s^{\text{eff.}}$ extraction, resulting in a total of 245 experimental data points used in our analysis. %
A more detailed discussion of the fitting procedure and a comparison to additional experimental results will be presented in a future long paper \cite{Faraday:2025}.

\begin{figure}[!th]
	\includegraphics[width=\linewidth]{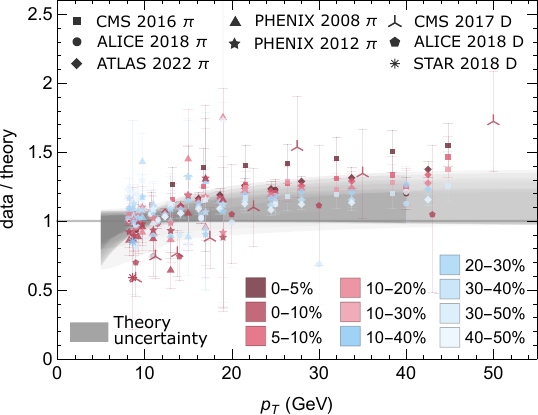}
	\caption{Measured $R_{AA}$ for various hadronic final states divided by the theoretical model expectation as a function of the transverse momentum $p_T$. Consult the text for details on data inclusion criteria. Statistical and systematic uncertainties are represented by error bars, and are added in quadrature for purely visual purposes. An estimate of the theoretical uncertainty is represented by the gray band around unity; see text for details.}
	\label{fig:data_over_theory_all_pions}
\end{figure}

\Cref{fig:data_over_theory_all_pions} shows a summary of results from the fitting procedure on central heavy-ion high-$p_T$ suppression data. 
The relative theoretical uncertainty is represented in the figure by a set of gray bands around unity. Specifically, each variation of our energy loss model---with a different cutoff $\kappa$ on the transverse radiated gluon momentum and a different collision energy loss model---is used to extract an independent value of $\alpha_s^{\text{eff.}}$ through the $\chi^2$ minimization procedure described above. The line at unity represents the theoretical prediction for $\kappa = 1$ and with HTL collisional energy loss, and the bands represent the theoretical predictions from other variations of the model. A different band is generated for each collision system as, in principle, each variation of the model may affect different datasets differently; however, the figure clearly shows a similar magnitude and $p_T$ dependence for the theoretical uncertainty.
The RHIC and LHC data are fit separately, resulting in a (potentially) different $\alpha_s^{\text{eff.}}$ at RHIC vs.\ LHC for each theoretical model, which is motivated by potential running coupling effects as a function of the temperature leading to different $\alpha_s^{\text{eff.}}$ at RHIC compared to LHC. 
We find that the best fit of $\alpha_s^{\text{eff.}}$ is always larger or similar at RHIC compared to LHC for all theoretical models, consistent with this hypothesis. 
Rather than examining the specific dependencies of our model, this figure highlights its overall agreement with a broad set of data. Future work \cite{Faraday:2025} will consider the detailed dependence of our results on flavor, $p_T$, centrality, and $\sqrt{s_{NN}}$.

\section{Qualitative, model-agnostic analysis}
\label{sec:model_agnostic}

In \cref{fig:small_vs_peripheral}, we saw that our convolved radiative and collisional energy loss model predicts nearly identical suppression in peripheral \coll{Au}{Au} and \coll{Pb}{Pb} collisions as in central \coll{d}{Au} and \coll{p}{Pb} collisions. A natural question to ask is: is the equal suppression in central small systems and peripheral large systems a feature which is unique to our energy loss model, or is it a robust feature of partonic energy loss in quark-gluon plasmas? In this section, we will consider the length-temperature phase space of central small, peripheral large, and central large collision systems, and examine the extent to which our conclusions are impacted by a different choice of energy loss model.

We consider energy loss of the form $\Delta E  \propto L^{a} T^{b} f(E)$ where $L$ is the path length, $T$ is the temperature, and $E$ is the incident energy. This form captures most simple, parametric forms of energy loss \cite{Horowitz:2011gd} including collisional energy loss
(collisional only) \cite{Bjorken:1982tu,Thoma:1990fm,Braaten:1991we}, Gyulassy-Levai-Vitev (GLV) radiative energy loss \cite{Gyulassy:2000er} (thin plasma limit, radiative only), BDMPS-Z radiative energy loss\footnote{We apply two limiting cases (high and low energy) of the BDMPS-Z formula over the entire $p_T$-range. This overestimates the theoretical uncertainty.} (thick plasma limit, radiative only) \cite{Baier:2001yt}, and AdS / CFT falling string energy loss (strong coupling limit) \cite{Chesler:2010bi,Marquet:2009eq}, where the parameters are shown in \cref{tab:parameters}. We take the Debye mass as a constant, $\mu_D = 0.5 ~\mathrm{GeV}$, but our results are insensitive to the temperature dependence. We include Bjorken expansion and approximate $R_{AB} \simeq 1 - n(p_T) \Delta E / E$, where $n(p_T) \equiv  - d \log( d \sigma) / d \log (p_T)$ is the spectral index, and $d \sigma$ is the production spectrum\footnote{This is valid for small energy loss and slowly varying $n(p_T)$ \cite{Faraday:2024gzx}.}\textsuperscript{,}\footnote{We consider only gluons and neglect hadronization to pions, but we checked that using only light quarks does not change our conclusions.}, which leads to $R_{AB} \simeq 1 - \beta n(p_T) L^{a - b / 3} T_0^b f(E) / E$, where $\beta$ is a proportionality constant.

\begin{table}[!h]
    \centering
    \begin{tabular}{lccc}
        \toprule
        Energy loss mechanism & $a$ & $b$ & $f (E)$ \\
        \midrule
        Collisional   & 1 & 2     & $\log \sfrac{E}{\mu_D}$ \\
        GLV & 2 & 3     & $\log \sfrac{E}{\mu_D}$ \\
				BDMPS-Z ($ E < \hat{q} L^2$)     & 1 & \(3/2\) & $E^{1 / 2}$ \\
				BDMPS-Z ($ E > \hat{q} L^2$)     & 2 & \(3\) & $1$\\
        AdS/CFT   & 3 & \(11/3\) & $E^{1 / 3}$ \\
        \bottomrule
    \end{tabular}
		\caption{Parameters corresponding to different energy loss models, where energy loss has the form \cite{Horowitz:2011gd} $\Delta E \propto L^a T^b f(E)$ with $L$ the path length, $T$ the temperature, $E$ is incident energy, and $\mu_D = 0.5 ~\mathrm{GeV}$ is the Debye mass.}
		\label{tab:parameters}
\end{table}

\begin{figure}[!b]
	\includegraphics[width=\linewidth]{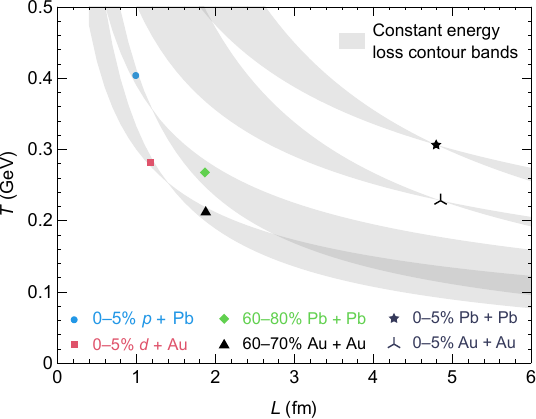}
	\caption{Average temperature $T$ (vertical axis) and length $L$ (horizontal axis) for 
		various collision systems.
	Bands formed from contours of constant energy loss using a variety of energy loss models are shown in gray.}
	\label{fig:length_temperature_phase_space}
\end{figure}

\Cref{fig:length_temperature_phase_space} plots the phase space of average path length $L$ and average temperature $T$ for $0\text{--}5\%$ centrality \coll{p}{Pb} collisions, $0\text{--}5\%$ centrality \coll{d}{Au} collisions, $60\text{--}80\%$ centrality \coll{Pb}{Pb} collisions, $60\text{--}70\%$ centrality \coll{Au}{Au} collisions, $0\text{--}5\%$ centrality \coll{Pb}{Pb} collisions, and $0\text{--}5\%$ centrality \coll{Au}{Au} collisions. 
The different parametric energy loss models from \cref{tab:parameters} are used to generate constant energy loss contours via $\Delta E / E = T^a L^b f(E = 10 ~\mathrm{GeV}) = \text{constant}$. The range of contours produced by the different energy loss models form a band of constant energy loss, which is constrained to pass through the collision systems of interest.

In \cref{fig:length_temperature_phase_space}, we observe that central small and peripheral large systems have $2\text{--}4$ times smaller lengths $L$ than the central heavy-ion collisions. 
	Additionally, central small systems are $\mathcal{O}(50\%)$ smaller and $\mathcal{O}(30 \text{--}50)\%$ hotter than their peripheral large system counterparts. 
The higher temperature in small systems results from shorter paths---leading to more time in the early-time high-temperature region---and from temperature fluctuations, rather than geometry, driving event activity. 
Despite the differences in energy loss models (\cref{tab:parameters}) and the non-negligible distance in phase space between peripheral large and central small systems, the constant energy loss contour bands in \cref{fig:length_temperature_phase_space} show that similar suppression is expected \emph{regardless} of the underlying energy loss model.
This can be understood qualitatively by considering two energy loss models with similar energy dependence: $\Delta E_{a} \sim L^a T^{a+1}$ and $\Delta E_{b} \sim L^b T^{b+1}$. Taylor expanding the ratio of the two about a point $(L^*, T^*)$ in phase space, where the models are constrained to have identical energy loss, yields $|1-\Delta E_{a} / \Delta E_{b}| \simeq  |(a-b) [ (L - L^*) / L^* + (T - T^*) / T^*]|$. Thus, the similarity of energy loss in peripheral large and central small systems is qualitatively understood as a combination of two factors:  the similarity of popular energy loss theories, $|a - b| \leq 2$, and the compensating effects of decreasing path length $L$ and increasing temperature $T$, [$(L - L^*) / L^* + (T - T^*) / T^* \lesssim 10\%$]. To quantify the impact of different energy loss models, we revisit the data-theory comparison with the various parametric energy loss models.

\begin{figure}[!t]
	\includegraphics[width=\linewidth]{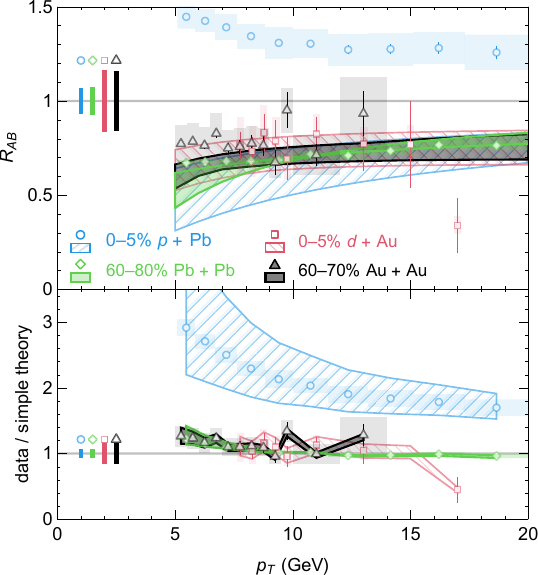}
	\caption{(top) Measured nuclear modification factor $R_{AB}$ (points) and simple theoretical model results (bands) as a function of $p_T$ for charged hadrons produced in $0\text{--}5\%$ centrality \coll{p}{Pb} and $60\text{--}80\%$ centrality \coll{Pb}{Pb} collisions at $\sqrt{s_{NN}} = 5.02 ~\mathrm{TeV}$ at LHC \cite{ATLAS:2022kqu}, as well as $0\text{--}5\%$ centrality \coll{d}{Au} \cite{PHENIX:2023dxl} and $60\text{--}70\%$ centrality \coll{Au}{Au} \cite{PHENIX:2012jha} collisions at $\sqrt{s_{NN}} = 0.2 ~\mathrm{TeV}$ at RHIC. (bottom) Measured $R_{AB}$ divided by simple theoretical model as a function of $p_T$ for the same systems as a function of $p_T$. Data divided by theory points are placed in the center of the theoretical uncertainty band.
	Statistical uncertainties are represented by square brackets, systematic uncertainties by shaded squares, global normalization uncertainties by the solid rectangles at unity, and theoretical uncertainties by the shaded bands. Points with statistical uncertainty $> 30 \%$ are excluded.}
	\label{fig:small_vs_peripheral_simple}
\end{figure}

\Cref{fig:small_vs_peripheral_simple} plots the experimentally measured $R_{AB}$ data (points) as well as the simple energy loss model results (bands) as a function of $p_T$, where the proportionality constant $\beta$ is constrained by a best fit to the $60\text{--}80\%$ centrality \coll{Pb}{Pb} data \cite{ATLAS:2022kqu}. The bottom panel shows the data divided by simple theory result.
Theoretical uncertainty bands represent the curves traced out by the various simple theoretical energy loss models.
The figure shows that the peripheral \coll{Au}{Au}, peripheral \coll{Pb}{Pb}, and central \coll{d}{Au} data are all in very good agreement with the theoretical expectation, while the central \coll{p}{Pb} data are inconsistent with all simple theoretical models. 
Additionally, the predicted $R_{AB}$ leads us to conclude that one expects similar suppression in central small and peripheral large systems \emph{regardless} of the underlying energy loss mechanism.
	Consistent with our analysis based on the $LT$ phase space, the theoretical uncertainty on the central \coll{d}{Au} and peripheral \coll{Au}{Au} points is $\lesssim 15\%$, even though this includes significantly different pictures for the underlying energy loss mechanism. The theoretical uncertainty on the central \coll{p}{Pb} points is larger, $\mathcal{O}(10\text{--}30)\%$; however, the \coll{p}{Pb} data points remain in marked disagreement with all of the simple energy loss models. The large uncertainty on the \coll{p}{Pb} points is because the $0\text{--}5\%$ \coll{p}{Pb} collision system is the farthest system from the fitted $60\text{--}80\%$ \coll{Pb}{Pb} system in the length-temperature phase space (see \cref{fig:length_temperature_phase_space}), which leads to it being significantly more sensitive to the different temperature and length power laws. 

\section{Summary and Outlook}
\label{sec:summary}

In this letter, we compared our small system size corrected energy loss model tuned to central large system high-$p_T$ suppression data to central small and peripheral large system high-$p_T$ suppression data. We found that for measurements with little sensitivity to centrality bias our model agreed well with data. 
For our model, we empirically saw equal suppression in central small systems and peripheral large systems. We then showed that this prediction of equal peripheral \coll{A}{A} and central \collThree{p}{d}{A} suppression is robust for a large class of energy loss models. 

In particular, we presented novel quantitative predictions for suppression in $\sqrt{s_{NN}} = 0.2 ~\mathrm{TeV}$ and $\sqrt{s_{NN}} = 5.02 ~\mathrm{TeV}$ central \collThree{p}{d}{A} and peripheral \coll{A}{A} collisions using our pQCD-based energy loss model tuned on central $\sqrt{s_{NN}} = 0.2 ~\mathrm{TeV}$ and $\sqrt{s_{NN}} = 5.02 ~\mathrm{TeV}$ \coll{A}{A} collision data. We included short path length corrections to both the radiative and collisional energy loss, which allowed for the first justifiable extrapolation of any energy loss model to small systems. Without additional tuning, our model reproduced experimental peripheral $\sqrt{s_{NN}} = 5.02 ~\mathrm{TeV}$ $R_{\text{PbPb}}$, peripheral $\sqrt{s_{NN}} = 0.2 ~\mathrm{TeV}$ $R_{\text{Au}\text{Au}}$, and central electroweak-normalized $\sqrt{s_{NN}} = 0.2 ~\mathrm{TeV}$ $R_{d \text{Au}}$ data \cite{ATLAS:2022kqu, PHENIX:2012jha,PHENIX:2023dxl}; however, our model predicted significant $\gtrsim 30\%$ suppression in central $\sqrt{s_{NN}} = 5.02 ~\mathrm{TeV}$ \coll{p}{Pb} collisions, in marked disagreement with the Glauber-normalized measured $\sim 20\%$ enhancement of $R_{p\text{Pb}}$ \cite{ATLAS:2022kqu}.

Peripheral \coll{A}{A} collisions involve $\sim \! 5$ times as many binary collisions as central \collThree{p}{d}{A} collisions, thus centrality bias---including various event selection biases and Glauber Model dependencies---should be more pronounced in central \collThree{p}{d}{A} compared to peripheral \coll{A}{A} collisions \cite{ALICE:2014xsp,ALICE:2018ekf}.
For this reason, and because we showed that for very general energy loss theories central $R_{p \text{Pb}} \sim $ peripheral $R_{AA}$, we expect that a significant source of the disagreement between our prediction in central \coll{p}{Pb} collisions and the measured $R_{pPb}$ is due to centrality bias.

Future experimental measurements which minimize centrality bias, such as $R_{pA}$ normalized by photons or $Z$ bosons \cite{PHENIX:2023dxl, Huss:2020dwe}, centrality averaged $R_{AA}$ measurements in \coll{O}{O} collisions \cite{Huss:2020dwe,Huss:2020whe,Citron:2018lsq}, and self-normalized jet substructure observables \cite{Kolbe:2023rsq} will help clarify the role of energy loss in small systems. 
Valuable future theoretical work in understanding both small and large system energy loss includes quantifying pre-thermalization time energy loss \cite{Zakharov:2008uk,Barata:2024xwy,Ilic:2021ezl}, relaxing the large formation time approximation \cite{Kolbe:2015rvk,Faraday:2023mmx,Wicks:2008zz,Faraday:2024gzx,Faraday:2023uay}, determining the scales at which the couplings run \cite{Horowitz:2010yg}, and improving our understanding of event selection biases in small systems \cite{JETSCAPE:2024dgu}.

\section*{Acknowledgments}

We thank Raymond Ehlers and Joseph Bahder for discussions. Computations were performed using facilities provided by the University of Cape Town’s ICTS High Performance Computing team: \href{http://hpc.uct.ac.za}{hpc.uct.ac.za}.
CF and WAH thank the National Research Foundation and the SA-CERN collaboration for their generous financial support during this work. 

\bibliographystyle{model1a-num-names} %
\bibliography{manual,letter_small_and_peripheral}

\begin{thebibliography}{71}
\expandafter\ifx\csname natexlab\endcsname\relax\def\natexlab#1{#1}\fi
\providecommand{\url}[1]{\texttt{#1}}
\providecommand{\href}[2]{#2}
\providecommand{\path}[1]{#1}
\providecommand{\DOIprefix}{doi:}
\providecommand{\ArXivprefix}{arXiv:}
\providecommand{\URLprefix}{URL: }
\providecommand{\Pubmedprefix}{pmid:}
\providecommand{\doi}[1]{\href{http://dx.doi.org/#1}{\path{#1}}}
\providecommand{\Pubmed}[1]{\href{pmid:#1}{\path{#1}}}
\providecommand{\bibinfo}[2]{#2}
\ifx\xfnm\relax \def\xfnm[#1]{\unskip,\space#1}\fi
%Type = Article
\bibitem[{Romatschke and Romatschke(2007)}]{Romatschke:2007mq}
\bibinfo{author}{P.~Romatschke}, \bibinfo{author}{U.~Romatschke}, \bibinfo{journal}{Phys. Rev. Lett.} \bibinfo{volume}{99} (\bibinfo{year}{2007}) \bibinfo{pages}{172301}. \DOIprefix\doi{10.1103/PhysRevLett.99.172301}. \href{http://arxiv.org/abs/0706.1522}{\tt arXiv:0706.1522}.
%Type = Article
\bibitem[{Song and Heinz(2008)}]{Song:2007ux}
\bibinfo{author}{H.~Song}, \bibinfo{author}{U.~W. Heinz}, \bibinfo{journal}{Phys. Rev. C} \bibinfo{volume}{77} (\bibinfo{year}{2008}) \bibinfo{pages}{064901}. \DOIprefix\doi{10.1103/PhysRevC.77.064901}. \href{http://arxiv.org/abs/0712.3715}{\tt arXiv:0712.3715}.
%Type = Article
\bibitem[{Schenke et~al.(2010)Schenke, Jeon, and Gale}]{Schenke:2010nt}
\bibinfo{author}{B.~Schenke}, \bibinfo{author}{S.~Jeon}, \bibinfo{author}{C.~Gale}, \bibinfo{journal}{Phys. Rev. C} \bibinfo{volume}{82} (\bibinfo{year}{2010}) \bibinfo{pages}{014903}. \DOIprefix\doi{10.1103/PhysRevC.82.014903}. \href{http://arxiv.org/abs/1004.1408}{\tt arXiv:1004.1408}.
%Type = Article
\bibitem[{Dainese et~al.(2005)Dainese, Loizides, and Paic}]{Dainese:2004te}
\bibinfo{author}{A.~Dainese}, \bibinfo{author}{C.~Loizides}, \bibinfo{author}{G.~Paic}, \bibinfo{journal}{Eur. Phys. J. C} \bibinfo{volume}{38} (\bibinfo{year}{2005}) \bibinfo{pages}{461--474}. \DOIprefix\doi{10.1140/epjc/s2004-02077-x}. \href{http://arxiv.org/abs/hep-ph/0406201}{\tt arXiv:hep-ph/0406201}.
%Type = Article
\bibitem[{Schenke et~al.(2009)Schenke, Gale, and Jeon}]{Schenke:2009gb}
\bibinfo{author}{B.~Schenke}, \bibinfo{author}{C.~Gale}, \bibinfo{author}{S.~Jeon}, \bibinfo{journal}{Phys. Rev. C} \bibinfo{volume}{80} (\bibinfo{year}{2009}) \bibinfo{pages}{054913}. \DOIprefix\doi{10.1103/PhysRevC.80.054913}. \href{http://arxiv.org/abs/0909.2037}{\tt arXiv:0909.2037}.
%Type = Article
\bibitem[{Wicks et~al.(2007)Wicks, Horowitz, Djordjevic, and Gyulassy}]{Wicks:2005gt}
\bibinfo{author}{S.~Wicks}, \bibinfo{author}{W.~Horowitz}, \bibinfo{author}{M.~Djordjevic}, \bibinfo{author}{M.~Gyulassy}, \bibinfo{journal}{Nucl. Phys. A} \bibinfo{volume}{784} (\bibinfo{year}{2007}) \bibinfo{pages}{426--442}. \DOIprefix\doi{10.1016/j.nuclphysa.2006.12.048}. \href{http://arxiv.org/abs/nucl-th/0512076}{\tt arXiv:nucl-th/0512076}.
%Type = Article
\bibitem[{Horowitz(2013)}]{Horowitz:2012cf}
\bibinfo{author}{W.~A. Horowitz}, \bibinfo{journal}{Nucl. Phys. A} \bibinfo{volume}{904-905} (\bibinfo{year}{2013}) \bibinfo{pages}{186c--193c}. \DOIprefix\doi{10.1016/j.nuclphysa.2013.01.061}. \href{http://arxiv.org/abs/1210.8330}{\tt arXiv:1210.8330}.
%Type = Article
\bibitem[{Adcox et~al.(2002)}]{PHENIX:2001hpc}
\bibinfo{author}{K.~Adcox}, et~al. (\bibinfo{collaboration}{PHENIX}), \bibinfo{journal}{Phys. Rev. Lett.} \bibinfo{volume}{88} (\bibinfo{year}{2002}) \bibinfo{pages}{022301}. \DOIprefix\doi{10.1103/PhysRevLett.88.022301}. \href{http://arxiv.org/abs/nucl-ex/0109003}{\tt arXiv:nucl-ex/0109003}.
%Type = Article
\bibitem[{Grosse-Oetringhaus and Wiedemann(2024)}]{Grosse-Oetringhaus:2024bwr}
\bibinfo{author}{J.~F. Grosse-Oetringhaus}, \bibinfo{author}{U.~A. Wiedemann}  (\bibinfo{year}{2024}). \href{http://arxiv.org/abs/2407.07484}{\tt arXiv:2407.07484}.
%Type = Article
\bibitem[{Aad et~al.(2013{\natexlab{a}})}]{ATLAS:2012cix}
\bibinfo{author}{G.~Aad}, et~al. (\bibinfo{collaboration}{ATLAS}), \bibinfo{journal}{Phys. Rev. Lett.} \bibinfo{volume}{110} (\bibinfo{year}{2013}{\natexlab{a}}) \bibinfo{pages}{182302}. \DOIprefix\doi{10.1103/PhysRevLett.110.182302}. \href{http://arxiv.org/abs/1212.5198}{\tt arXiv:1212.5198}.
%Type = Article
\bibitem[{Aad et~al.(2013{\natexlab{b}})}]{ATLAS:2013jmi}
\bibinfo{author}{G.~Aad}, et~al. (\bibinfo{collaboration}{ATLAS}), \bibinfo{journal}{Phys. Lett. B} \bibinfo{volume}{725} (\bibinfo{year}{2013}{\natexlab{b}}) \bibinfo{pages}{60--78}. \DOIprefix\doi{10.1016/j.physletb.2013.06.057}. \href{http://arxiv.org/abs/1303.2084}{\tt arXiv:1303.2084}.
%Type = Article
\bibitem[{Abelev et~al.(2014)}]{ALICE:2014dwt}
\bibinfo{author}{B.~B. Abelev}, et~al. (\bibinfo{collaboration}{ALICE}), \bibinfo{journal}{Phys. Rev. C} \bibinfo{volume}{90} (\bibinfo{year}{2014}) \bibinfo{pages}{054901}. \DOIprefix\doi{10.1103/PhysRevC.90.054901}. \href{http://arxiv.org/abs/1406.2474}{\tt arXiv:1406.2474}.
%Type = Article
\bibitem[{Khachatryan et~al.(2015)}]{CMS:2015yux}
\bibinfo{author}{V.~Khachatryan}, et~al. (\bibinfo{collaboration}{CMS}), \bibinfo{journal}{Phys. Rev. Lett.} \bibinfo{volume}{115} (\bibinfo{year}{2015}) \bibinfo{pages}{012301}. \DOIprefix\doi{10.1103/PhysRevLett.115.012301}. \href{http://arxiv.org/abs/1502.05382}{\tt arXiv:1502.05382}.
%Type = Article
\bibitem[{Adam et~al.(2016)}]{ALICE:2015mpp}
\bibinfo{author}{J.~Adam}, et~al. (\bibinfo{collaboration}{ALICE}), \bibinfo{journal}{Phys. Lett. B} \bibinfo{volume}{758} (\bibinfo{year}{2016}) \bibinfo{pages}{389--401}. \DOIprefix\doi{10.1016/j.physletb.2016.05.027}. \href{http://arxiv.org/abs/1512.07227}{\tt arXiv:1512.07227}.
%Type = Article
\bibitem[{Abelev et~al.(2014)}]{ALICE:2013wgn}
\bibinfo{author}{B.~B. Abelev}, et~al. (\bibinfo{collaboration}{ALICE}), \bibinfo{journal}{Phys. Lett. B} \bibinfo{volume}{728} (\bibinfo{year}{2014}) \bibinfo{pages}{25--38}. \DOIprefix\doi{10.1016/j.physletb.2013.11.020}. \href{http://arxiv.org/abs/1307.6796}{\tt arXiv:1307.6796}.
%Type = Article
\bibitem[{Adam et~al.(2016)}]{ALICE:2016sdt}
\bibinfo{author}{J.~Adam}, et~al. (\bibinfo{collaboration}{ALICE}), \bibinfo{journal}{JHEP} \bibinfo{volume}{06} (\bibinfo{year}{2016}) \bibinfo{pages}{050}. \DOIprefix\doi{10.1007/JHEP06(2016)050}. \href{http://arxiv.org/abs/1603.02816}{\tt arXiv:1603.02816}.
%Type = Article
\bibitem[{Aidala et~al.(2019)}]{PHENIX:2018lia}
\bibinfo{author}{C.~Aidala}, et~al. (\bibinfo{collaboration}{PHENIX}), \bibinfo{journal}{Nature Phys.} \bibinfo{volume}{15} (\bibinfo{year}{2019}) \bibinfo{pages}{214--220}. \DOIprefix\doi{10.1038/s41567-018-0360-0}. \href{http://arxiv.org/abs/1805.02973}{\tt arXiv:1805.02973}.
%Type = Article
\bibitem[{Weller and Romatschke(2017)}]{Weller:2017tsr}
\bibinfo{author}{R.~D. Weller}, \bibinfo{author}{P.~Romatschke}, \bibinfo{journal}{Phys. Lett. B} \bibinfo{volume}{774} (\bibinfo{year}{2017}) \bibinfo{pages}{351--356}. \DOIprefix\doi{10.1016/j.physletb.2017.09.077}. \href{http://arxiv.org/abs/1701.07145}{\tt arXiv:1701.07145}.
%Type = Article
\bibitem[{Schenke et~al.(2020)Schenke, Shen, and Tribedy}]{Schenke:2020mbo}
\bibinfo{author}{B.~Schenke}, \bibinfo{author}{C.~Shen}, \bibinfo{author}{P.~Tribedy}, \bibinfo{journal}{Phys. Rev. C} \bibinfo{volume}{102} (\bibinfo{year}{2020}) \bibinfo{pages}{044905}. \DOIprefix\doi{10.1103/PhysRevC.102.044905}. \href{http://arxiv.org/abs/2005.14682}{\tt arXiv:2005.14682}.
%Type = Article
\bibitem[{Aad et~al.(2023)}]{ATLAS:2022kqu}
\bibinfo{author}{G.~Aad}, et~al. (\bibinfo{collaboration}{ATLAS}), \bibinfo{journal}{JHEP} \bibinfo{volume}{07} (\bibinfo{year}{2023}) \bibinfo{pages}{074}. \DOIprefix\doi{10.1007/JHEP07(2023)074}. \href{http://arxiv.org/abs/2211.15257}{\tt arXiv:2211.15257}.
%Type = Article
\bibitem[{Abdulameer et~al.(2025)}]{PHENIX:2023dxl}
\bibinfo{author}{N.~J. Abdulameer}, et~al. (\bibinfo{collaboration}{PHENIX}), \bibinfo{journal}{Phys. Rev. Lett.} \bibinfo{volume}{134} (\bibinfo{year}{2025}) \bibinfo{pages}{022302}. \DOIprefix\doi{10.1103/PhysRevLett.134.022302}. \href{http://arxiv.org/abs/2303.12899}{\tt arXiv:2303.12899}.
%Type = Article
\bibitem[{Adam et~al.(2015)}]{ALICE:2014xsp}
\bibinfo{author}{J.~Adam}, et~al. (\bibinfo{collaboration}{ALICE}), \bibinfo{journal}{Phys. Rev. C} \bibinfo{volume}{91} (\bibinfo{year}{2015}) \bibinfo{pages}{064905}. \DOIprefix\doi{10.1103/PhysRevC.91.064905}. \href{http://arxiv.org/abs/1412.6828}{\tt arXiv:1412.6828}.
%Type = Article
\bibitem[{Acharya et~al.(2019)}]{ALICE:2018ekf}
\bibinfo{author}{S.~Acharya}, et~al. (\bibinfo{collaboration}{ALICE}), \bibinfo{journal}{Phys. Lett. B} \bibinfo{volume}{793} (\bibinfo{year}{2019}) \bibinfo{pages}{420--432}. \DOIprefix\doi{10.1016/j.physletb.2019.04.047}. \href{http://arxiv.org/abs/1805.05212}{\tt arXiv:1805.05212}.
%Type = Article
\bibitem[{Aad et~al.(2016)}]{ATLAS:2016xpn}
\bibinfo{author}{G.~Aad}, et~al. (\bibinfo{collaboration}{ATLAS}), \bibinfo{journal}{Phys. Lett. B} \bibinfo{volume}{763} (\bibinfo{year}{2016}) \bibinfo{pages}{313--336}. \DOIprefix\doi{10.1016/j.physletb.2016.10.053}. \href{http://arxiv.org/abs/1605.06436}{\tt arXiv:1605.06436}.
%Type = Article
\bibitem[{Miller et~al.(2007)Miller, Reygers, Sanders, and Steinberg}]{Miller:2007ri}
\bibinfo{author}{M.~L. Miller}, \bibinfo{author}{K.~Reygers}, \bibinfo{author}{S.~J. Sanders}, \bibinfo{author}{P.~Steinberg}, \bibinfo{journal}{Ann. Rev. Nucl. Part. Sci.} \bibinfo{volume}{57} (\bibinfo{year}{2007}) \bibinfo{pages}{205--243}. \DOIprefix\doi{10.1146/annurev.nucl.57.090506.123020}. \href{http://arxiv.org/abs/nucl-ex/0701025}{\tt arXiv:nucl-ex/0701025}.
%Type = Article
\bibitem[{Faraday et~al.(2023)Faraday, Grindrod, and Horowitz}]{Faraday:2023mmx}
\bibinfo{author}{C.~Faraday}, \bibinfo{author}{A.~Grindrod}, \bibinfo{author}{W.~A. Horowitz}, \bibinfo{journal}{Eur. Phys. J. C} \bibinfo{volume}{83} (\bibinfo{year}{2023}) \bibinfo{pages}{1060}. \DOIprefix\doi{10.1140/epjc/s10052-023-12234-y}. \href{http://arxiv.org/abs/2305.13182}{\tt arXiv:2305.13182}.
%Type = Article
\bibitem[{Kolbe and Horowitz(2019)}]{Kolbe:2015rvk}
\bibinfo{author}{I.~Kolbe}, \bibinfo{author}{W.~A. Horowitz}, \bibinfo{journal}{Phys. Rev. C} \bibinfo{volume}{100} (\bibinfo{year}{2019}) \bibinfo{pages}{024913}. \DOIprefix\doi{10.1103/PhysRevC.100.024913}. \href{http://arxiv.org/abs/1511.09313}{\tt arXiv:1511.09313}.
%Type = Article
\bibitem[{Faraday and Horowitz(2024)}]{Faraday:2024gzx}
\bibinfo{author}{C.~Faraday}, \bibinfo{author}{W.~A. Horowitz}  (\bibinfo{year}{2024}). \href{http://arxiv.org/abs/2408.14426}{\tt arXiv:2408.14426}.
%Type = Phdthesis
\bibitem[{Wicks(2008)}]{Wicks:2008zz}
\bibinfo{author}{S.~Wicks}, \bibinfo{type}{{PhD} thesis}, \bibinfo{year}{2008}.
%Type = Article
\bibitem[{Ehlers et~al.(2024)}]{JETSCAPE:2024cqe}
\bibinfo{author}{R.~Ehlers}, et~al. (\bibinfo{collaboration}{JETSCAPE})  (\bibinfo{year}{2024}). \href{http://arxiv.org/abs/2408.08247}{\tt arXiv:2408.08247}.
%Type = Article
\bibitem[{Eskola et~al.(2022)Eskola, Paakkinen, Paukkunen, and Salgado}]{Eskola:2021nhw}
\bibinfo{author}{K.~J. Eskola}, \bibinfo{author}{P.~Paakkinen}, \bibinfo{author}{H.~Paukkunen}, \bibinfo{author}{C.~A. Salgado}, \bibinfo{journal}{Eur. Phys. J. C} \bibinfo{volume}{82} (\bibinfo{year}{2022}) \bibinfo{pages}{413}. \DOIprefix\doi{10.1140/epjc/s10052-022-10359-0}. \href{http://arxiv.org/abs/2112.12462}{\tt arXiv:2112.12462}.
%Type = Article
\bibitem[{Armesto et~al.(2022)Armesto, Lappi, M\"antysaari, Paukkunen, and Tevio}]{Armesto:2022mxy}
\bibinfo{author}{N.~Armesto}, \bibinfo{author}{T.~Lappi}, \bibinfo{author}{H.~M\"antysaari}, \bibinfo{author}{H.~Paukkunen}, \bibinfo{author}{M.~Tevio}, \bibinfo{journal}{Phys. Rev. D} \bibinfo{volume}{105} (\bibinfo{year}{2022}) \bibinfo{pages}{114017}. \DOIprefix\doi{10.1103/PhysRevD.105.114017}. \href{http://arxiv.org/abs/2203.05846}{\tt arXiv:2203.05846}.
%Type = Article
\bibitem[{Ke and Vitev(2023)}]{Ke:2022gkq}
\bibinfo{author}{W.~Ke}, \bibinfo{author}{I.~Vitev}, \bibinfo{journal}{Phys. Rev. C} \bibinfo{volume}{107} (\bibinfo{year}{2023}) \bibinfo{pages}{064903}. \DOIprefix\doi{10.1103/PhysRevC.107.064903}. \href{http://arxiv.org/abs/2204.00634}{\tt arXiv:2204.00634}.
%Type = Article
\bibitem[{Perepelitsa(2024)}]{Perepelitsa:2024eik}
\bibinfo{author}{D.~V. Perepelitsa}, \bibinfo{journal}{Phys. Rev. C} \bibinfo{volume}{110} (\bibinfo{year}{2024}) \bibinfo{pages}{L011901}. \DOIprefix\doi{10.1103/PhysRevC.110.L011901}. \href{http://arxiv.org/abs/2404.17660}{\tt arXiv:2404.17660}.
%Type = Article
\bibitem[{Fries et~al.(2008)Fries, Greco, and Sorensen}]{Fries:2008hs}
\bibinfo{author}{R.~J. Fries}, \bibinfo{author}{V.~Greco}, \bibinfo{author}{P.~Sorensen}, \bibinfo{journal}{Ann. Rev. Nucl. Part. Sci.} \bibinfo{volume}{58} (\bibinfo{year}{2008}) \bibinfo{pages}{177--205}. \DOIprefix\doi{10.1146/annurev.nucl.58.110707.171134}. \href{http://arxiv.org/abs/0807.4939}{\tt arXiv:0807.4939}.
%Type = Article
\bibitem[{Zhao and Rapp(2011)}]{Zhao:2011cv}
\bibinfo{author}{X.~Zhao}, \bibinfo{author}{R.~Rapp}, \bibinfo{journal}{Nucl. Phys. A} \bibinfo{volume}{859} (\bibinfo{year}{2011}) \bibinfo{pages}{114--125}. \DOIprefix\doi{10.1016/j.nuclphysa.2011.05.001}. \href{http://arxiv.org/abs/1102.2194}{\tt arXiv:1102.2194}.
%Type = Article
\bibitem[{Adare et~al.(2013)}]{PHENIX:2012jha}
\bibinfo{author}{A.~Adare}, et~al. (\bibinfo{collaboration}{PHENIX}), \bibinfo{journal}{Phys. Rev. C} \bibinfo{volume}{87} (\bibinfo{year}{2013}) \bibinfo{pages}{034911}. \DOIprefix\doi{10.1103/PhysRevC.87.034911}. \href{http://arxiv.org/abs/1208.2254}{\tt arXiv:1208.2254}.
%Type = Article
\bibitem[{Soudi et~al.(2024)}]{JETSCAPE:2024dgu}
\bibinfo{author}{I.~Soudi}, et~al. (\bibinfo{collaboration}{JETSCAPE})  (\bibinfo{year}{2024}). \href{http://arxiv.org/abs/2407.17443}{\tt arXiv:2407.17443}.
%Type = Article
\bibitem[{Cao et~al.(2021)}]{JETSCAPE:2021ehl}
\bibinfo{author}{S.~Cao}, et~al. (\bibinfo{collaboration}{JETSCAPE}), \bibinfo{journal}{Phys. Rev. C} \bibinfo{volume}{104} (\bibinfo{year}{2021}) \bibinfo{pages}{024905}. \DOIprefix\doi{10.1103/PhysRevC.104.024905}. \href{http://arxiv.org/abs/2102.11337}{\tt arXiv:2102.11337}.
%Type = Article
\bibitem[{Huss et~al.(2021{\natexlab{a}})Huss, Kurkela, Mazeliauskas, Paatelainen, van~der Schee, and Wiedemann}]{Huss:2020dwe}
\bibinfo{author}{A.~Huss}, \bibinfo{author}{A.~Kurkela}, \bibinfo{author}{A.~Mazeliauskas}, \bibinfo{author}{R.~Paatelainen}, \bibinfo{author}{W.~van~der Schee}, \bibinfo{author}{U.~A. Wiedemann}, \bibinfo{journal}{Phys. Rev. Lett.} \bibinfo{volume}{126} (\bibinfo{year}{2021}{\natexlab{a}}) \bibinfo{pages}{192301}. \DOIprefix\doi{10.1103/PhysRevLett.126.192301}. \href{http://arxiv.org/abs/2007.13754}{\tt arXiv:2007.13754}.
%Type = Article
\bibitem[{Huss et~al.(2021{\natexlab{b}})Huss, Kurkela, Mazeliauskas, Paatelainen, van~der Schee, and Wiedemann}]{Huss:2020whe}
\bibinfo{author}{A.~Huss}, \bibinfo{author}{A.~Kurkela}, \bibinfo{author}{A.~Mazeliauskas}, \bibinfo{author}{R.~Paatelainen}, \bibinfo{author}{W.~van~der Schee}, \bibinfo{author}{U.~A. Wiedemann}, \bibinfo{journal}{Phys. Rev. C} \bibinfo{volume}{103} (\bibinfo{year}{2021}{\natexlab{b}}) \bibinfo{pages}{054903}. \DOIprefix\doi{10.1103/PhysRevC.103.054903}. \href{http://arxiv.org/abs/2007.13758}{\tt arXiv:2007.13758}.
%Type = Article
\bibitem[{Zakharov(2021)}]{Zakharov:2021uza}
\bibinfo{author}{B.~G. Zakharov}, \bibinfo{journal}{JHEP} \bibinfo{volume}{09} (\bibinfo{year}{2021}) \bibinfo{pages}{087}. \DOIprefix\doi{10.1007/JHEP09(2021)087}. \href{http://arxiv.org/abs/2105.09350}{\tt arXiv:2105.09350}.
%Type = Article
\bibitem[{Djordjevic and Gyulassy(2004)}]{Djordjevic:2003zk}
\bibinfo{author}{M.~Djordjevic}, \bibinfo{author}{M.~Gyulassy}, \bibinfo{journal}{Nucl. Phys. A} \bibinfo{volume}{733} (\bibinfo{year}{2004}) \bibinfo{pages}{265--298}. \DOIprefix\doi{10.1016/j.nuclphysa.2003.12.020}. \href{http://arxiv.org/abs/nucl-th/0310076}{\tt arXiv:nucl-th/0310076}.
%Type = Article
\bibitem[{Braaten and Thoma(1991)}]{Braaten:1991we}
\bibinfo{author}{E.~Braaten}, \bibinfo{author}{M.~H. Thoma}, \bibinfo{journal}{Phys. Rev. D} \bibinfo{volume}{44} (\bibinfo{year}{1991}) \bibinfo{pages}{R2625}. \DOIprefix\doi{10.1103/PhysRevD.44.R2625}.
%Type = Inproceedings
\bibitem[{Faraday and Horowitz(2023)}]{Faraday:2023uay}
\bibinfo{author}{C.~Faraday}, \bibinfo{author}{W.~A. Horowitz}, in: \bibinfo{booktitle}{{67th Annual Conference of the South African Institute of Physics}}. \href{http://arxiv.org/abs/2309.06246}{\tt arXiv:2309.06246}.
%Type = Article
\bibitem[{Horowitz and Cole(2010)}]{Horowitz:2009eb}
\bibinfo{author}{W.~A. Horowitz}, \bibinfo{author}{B.~A. Cole}, \bibinfo{journal}{Phys. Rev. C} \bibinfo{volume}{81} (\bibinfo{year}{2010}) \bibinfo{pages}{024909}. \DOIprefix\doi{10.1103/PhysRevC.81.024909}. \href{http://arxiv.org/abs/0910.1823}{\tt arXiv:0910.1823}.
%Type = Article
\bibitem[{Armesto et~al.(2012)}]{Armesto:2011ht}
\bibinfo{author}{N.~Armesto}, et~al., \bibinfo{journal}{Phys. Rev. C} \bibinfo{volume}{86} (\bibinfo{year}{2012}) \bibinfo{pages}{064904}. \DOIprefix\doi{10.1103/PhysRevC.86.064904}. \href{http://arxiv.org/abs/1106.1106}{\tt arXiv:1106.1106}.
%Type = Article
\bibitem[{Xu et~al.(2014)Xu, Buzzatti, and Gyulassy}]{Xu:2014ica}
\bibinfo{author}{J.~Xu}, \bibinfo{author}{A.~Buzzatti}, \bibinfo{author}{M.~Gyulassy}, \bibinfo{journal}{JHEP} \bibinfo{volume}{08} (\bibinfo{year}{2014}) \bibinfo{pages}{063}. \DOIprefix\doi{10.1007/JHEP08(2014)063}. \href{http://arxiv.org/abs/1402.2956}{\tt arXiv:1402.2956}.
%Type = Article
\bibitem[{Djordjevic and Djordjevic(2014)}]{Djordjevic:2013xoa}
\bibinfo{author}{M.~Djordjevic}, \bibinfo{author}{M.~Djordjevic}, \bibinfo{journal}{Phys. Lett. B} \bibinfo{volume}{734} (\bibinfo{year}{2014}) \bibinfo{pages}{286--289}. \DOIprefix\doi{10.1016/j.physletb.2014.05.053}. \href{http://arxiv.org/abs/1307.4098}{\tt arXiv:1307.4098}.
%Type = Article
\bibitem[{Peshier(2006)}]{Peshier:2006ah}
\bibinfo{author}{A.~Peshier}  (\bibinfo{year}{2006}). \href{http://arxiv.org/abs/hep-ph/0601119}{\tt arXiv:hep-ph/0601119}.
%Type = Article
\bibitem[{Adare et~al.(2008)}]{PHENIX:2008ove}
\bibinfo{author}{A.~Adare}, et~al. (\bibinfo{collaboration}{PHENIX}), \bibinfo{journal}{Phys. Rev. C} \bibinfo{volume}{77} (\bibinfo{year}{2008}) \bibinfo{pages}{064907}. \DOIprefix\doi{10.1103/PhysRevC.77.064907}. \href{http://arxiv.org/abs/0801.1665}{\tt arXiv:0801.1665}.
%Type = Article
\bibitem[{Khachatryan et~al.(2017)}]{CMS:2016xef}
\bibinfo{author}{V.~Khachatryan}, et~al. (\bibinfo{collaboration}{CMS}), \bibinfo{journal}{JHEP} \bibinfo{volume}{04} (\bibinfo{year}{2017}) \bibinfo{pages}{039}. \DOIprefix\doi{10.1007/JHEP04(2017)039}. \href{http://arxiv.org/abs/1611.01664}{\tt arXiv:1611.01664}.
%Type = Article
\bibitem[{Acharya et~al.(2018)}]{ALICE:2018vuu}
\bibinfo{author}{S.~Acharya}, et~al. (\bibinfo{collaboration}{ALICE}), \bibinfo{journal}{JHEP} \bibinfo{volume}{11} (\bibinfo{year}{2018}) \bibinfo{pages}{013}. \DOIprefix\doi{10.1007/JHEP11(2018)013}. \href{http://arxiv.org/abs/1802.09145}{\tt arXiv:1802.09145}.
%Type = Article
\bibitem[{Sirunyan et~al.(2018)}]{CMS:2017qjw}
\bibinfo{author}{A.~M. Sirunyan}, et~al. (\bibinfo{collaboration}{CMS}), \bibinfo{journal}{Phys. Lett. B} \bibinfo{volume}{782} (\bibinfo{year}{2018}) \bibinfo{pages}{474--496}. \DOIprefix\doi{10.1016/j.physletb.2018.05.074}. \href{http://arxiv.org/abs/1708.04962}{\tt arXiv:1708.04962}.
%Type = Article
\bibitem[{Acharya et~al.(2018)}]{ALICE:2018lyv}
\bibinfo{author}{S.~Acharya}, et~al. (\bibinfo{collaboration}{ALICE}), \bibinfo{journal}{JHEP} \bibinfo{volume}{10} (\bibinfo{year}{2018}) \bibinfo{pages}{174}. \DOIprefix\doi{10.1007/JHEP10(2018)174}. \href{http://arxiv.org/abs/1804.09083}{\tt arXiv:1804.09083}.
%Type = Article
\bibitem[{Adare et~al.(2008)}]{PHENIX:2008saf}
\bibinfo{author}{A.~Adare}, et~al. (\bibinfo{collaboration}{PHENIX}), \bibinfo{journal}{Phys. Rev. Lett.} \bibinfo{volume}{101} (\bibinfo{year}{2008}) \bibinfo{pages}{232301}. \DOIprefix\doi{10.1103/PhysRevLett.101.232301}. \href{http://arxiv.org/abs/0801.4020}{\tt arXiv:0801.4020}.
%Type = Article
\bibitem[{Adam et~al.(2019)}]{STAR:2018zdy}
\bibinfo{author}{J.~Adam}, et~al. (\bibinfo{collaboration}{STAR}), \bibinfo{journal}{Phys. Rev. C} \bibinfo{volume}{99} (\bibinfo{year}{2019}) \bibinfo{pages}{034908}. \DOIprefix\doi{10.1103/PhysRevC.99.034908}. \href{http://arxiv.org/abs/1812.10224}{\tt arXiv:1812.10224}.
%Type = Unpublished
\bibitem[{Faraday and Horowitz(2025)}]{Faraday:2025}
\bibinfo{author}{C.~Faraday}, \bibinfo{author}{W.~A. Horowitz}, \bibinfo{year}{2025}. \bibinfo{note}{Work in preparation}.
%Type = Article
\bibitem[{Horowitz and Gyulassy(2011)}]{Horowitz:2011gd}
\bibinfo{author}{W.~A. Horowitz}, \bibinfo{author}{M.~Gyulassy}, \bibinfo{journal}{Nucl. Phys. A} \bibinfo{volume}{872} (\bibinfo{year}{2011}) \bibinfo{pages}{265--285}. \DOIprefix\doi{10.1016/j.nuclphysa.2011.09.018}. \href{http://arxiv.org/abs/1104.4958}{\tt arXiv:1104.4958}.
%Type = Article
\bibitem[{Bjorken(1982)}]{Bjorken:1982tu}
\bibinfo{author}{J.~D. Bjorken}  (\bibinfo{year}{1982}).
%Type = Article
\bibitem[{Thoma and Gyulassy(1991)}]{Thoma:1990fm}
\bibinfo{author}{M.~H. Thoma}, \bibinfo{author}{M.~Gyulassy}, \bibinfo{journal}{Nucl. Phys. B} \bibinfo{volume}{351} (\bibinfo{year}{1991}) \bibinfo{pages}{491--506}. \DOIprefix\doi{10.1016/S0550-3213(05)80031-8}.
%Type = Article
\bibitem[{Gyulassy et~al.(2001)Gyulassy, Levai, and Vitev}]{Gyulassy:2000er}
\bibinfo{author}{M.~Gyulassy}, \bibinfo{author}{P.~Levai}, \bibinfo{author}{I.~Vitev}, \bibinfo{journal}{Nucl. Phys. B} \bibinfo{volume}{594} (\bibinfo{year}{2001}) \bibinfo{pages}{371--419}. \DOIprefix\doi{10.1016/S0550-3213(00)00652-0}. \href{http://arxiv.org/abs/nucl-th/0006010}{\tt arXiv:nucl-th/0006010}.
%Type = Article
\bibitem[{Baier et~al.(2001)Baier, Dokshitzer, Mueller, and Schiff}]{Baier:2001yt}
\bibinfo{author}{R.~Baier}, \bibinfo{author}{Y.~L. Dokshitzer}, \bibinfo{author}{A.~H. Mueller}, \bibinfo{author}{D.~Schiff}, \bibinfo{journal}{JHEP} \bibinfo{volume}{09} (\bibinfo{year}{2001}) \bibinfo{pages}{033}. \DOIprefix\doi{10.1088/1126-6708/2001/09/033}. \href{http://arxiv.org/abs/hep-ph/0106347}{\tt arXiv:hep-ph/0106347}.
%Type = Article
\bibitem[{Chesler and Yaffe(2011)}]{Chesler:2010bi}
\bibinfo{author}{P.~M. Chesler}, \bibinfo{author}{L.~G. Yaffe}, \bibinfo{journal}{Phys. Rev. Lett.} \bibinfo{volume}{106} (\bibinfo{year}{2011}) \bibinfo{pages}{021601}. \DOIprefix\doi{10.1103/PhysRevLett.106.021601}. \href{http://arxiv.org/abs/1011.3562}{\tt arXiv:1011.3562}.
%Type = Article
\bibitem[{Marquet and Renk(2010)}]{Marquet:2009eq}
\bibinfo{author}{C.~Marquet}, \bibinfo{author}{T.~Renk}, \bibinfo{journal}{Phys. Lett. B} \bibinfo{volume}{685} (\bibinfo{year}{2010}) \bibinfo{pages}{270--276}. \DOIprefix\doi{10.1016/j.physletb.2010.01.076}. \href{http://arxiv.org/abs/0908.0880}{\tt arXiv:0908.0880}.
%Type = Article
\bibitem[{Citron et~al.(2019)}]{Citron:2018lsq}
\bibinfo{author}{Z.~Citron}, et~al., \bibinfo{journal}{CERN Yellow Rep. Monogr.} \bibinfo{volume}{7} (\bibinfo{year}{2019}) \bibinfo{pages}{1159--1410}. \DOIprefix\doi{10.23731/CYRM-2019-007.1159}. \href{http://arxiv.org/abs/1812.06772}{\tt arXiv:1812.06772}.
%Type = Article
\bibitem[{Kolb\'e(2023)}]{Kolbe:2023rsq}
\bibinfo{author}{I.~Kolb\'e}  (\bibinfo{year}{2023}). \href{http://arxiv.org/abs/2303.14166}{\tt arXiv:2303.14166}.
%Type = Article
\bibitem[{Zakharov(2008)}]{Zakharov:2008uk}
\bibinfo{author}{B.~G. Zakharov}, \bibinfo{journal}{JETP Lett.} \bibinfo{volume}{88} (\bibinfo{year}{2008}) \bibinfo{pages}{475--480}. \DOIprefix\doi{10.1134/S0021364008200022}. \href{http://arxiv.org/abs/0809.0599}{\tt arXiv:0809.0599}.
%Type = Article
\bibitem[{Barata et~al.(2024)Barata, Hauksson, Mayo~L\'opez, and Sadofyev}]{Barata:2024xwy}
\bibinfo{author}{J.~a. Barata}, \bibinfo{author}{S.~Hauksson}, \bibinfo{author}{X.~Mayo~L\'opez}, \bibinfo{author}{A.~V. Sadofyev}, \bibinfo{journal}{Phys. Rev. D} \bibinfo{volume}{110} (\bibinfo{year}{2024}) \bibinfo{pages}{094055}. \DOIprefix\doi{10.1103/PhysRevD.110.094055}. \href{http://arxiv.org/abs/2406.07615}{\tt arXiv:2406.07615}.
%Type = Article
\bibitem[{Ilic et~al.(2021)Ilic, Zigic, Djordjevic, and Djordjevic}]{Ilic:2021ezl}
\bibinfo{author}{B.~Ilic}, \bibinfo{author}{D.~Zigic}, \bibinfo{author}{M.~Djordjevic}, \bibinfo{author}{M.~Djordjevic}, \bibinfo{journal}{Int. J. Mod. Phys. E} \bibinfo{volume}{30} (\bibinfo{year}{2021}) \bibinfo{pages}{2141007}. \DOIprefix\doi{10.1142/S021830132141007X}. \href{http://arxiv.org/abs/2203.11883}{\tt arXiv:2203.11883}.
%Type = Article
\bibitem[{Horowitz and Kovchegov(2011)}]{Horowitz:2010yg}
\bibinfo{author}{W.~A. Horowitz}, \bibinfo{author}{Y.~V. Kovchegov}, \bibinfo{journal}{Nucl. Phys. A} \bibinfo{volume}{849} (\bibinfo{year}{2011}) \bibinfo{pages}{72--97}. \DOIprefix\doi{10.1016/j.nuclphysa.2010.10.014}. \href{http://arxiv.org/abs/1009.0545}{\tt arXiv:1009.0545}.

\end{thebibliography}

\end{document}